\documentclass[aps,prl,showpacs,amsmath,amssymb,twocolumn,superscriptaddress]{revtex4-1}
\usepackage{amsmath}
\usepackage{amsfonts}
\usepackage{amssymb}
\usepackage{graphicx}
\usepackage{epsfig}
\usepackage{color}
\usepackage{empheq}
\usepackage{epstopdf}
\newcommand{\RR}{\right}
\newcommand{\LL}{\left}
\newcommand{\m}{\mathrm}
\newcommand{\dg}{\dagger}

\newcommand{\fref}[1]{Fig.~\ref{#1}}

\newcommand{\vari}{}

\begin{document}
\title{Noiseless quantum measurement and squeezing of microwave fields utilizing mechanical vibrations}

\author{C. F. Ockeloen-Korppi}
\affiliation{Department of Applied Physics, Aalto University, P.O. Box 15100, FI-00076 AALTO, Finland}
\author{E. Damsk\"agg}
\affiliation{Department of Applied Physics, Aalto University, P.O. Box 15100, FI-00076 AALTO, Finland}
\author{J.-M. Pirkkalainen}
\affiliation{Department of Applied Physics, Aalto University, P.O. Box 15100, FI-00076 AALTO, Finland}
\author{T.~T.~Heikkil\"a}
\author{F. Massel}
\affiliation{Department of Physics and Nanoscience Center, University of Jyv\"askyl\"a,
P.O. Box 35 (YFL), FI-40014 University of Jyv\"askyl\"a, Finland}
\author{M.~A.~Sillanp\"a\"a}
\email[]{mika.sillanpaa@aalto.fi}
\affiliation{Department of Applied Physics, Aalto University, P.O. Box 15100, FI-00076 AALTO, Finland}
\begin{abstract}

A process which strongly amplifies both quadrature amplitudes of an oscillatory signal necessarily adds noise.  Alternatively, if the information in one quadrature is lost in phase-sensitive amplification, it is possible to completely reconstruct the other quadrature. Here we demonstrate such a nearly perfect phase-sensitive measurement using a cavity optomechanical scheme, characterized by an extremely small noise less than 0.2 quanta. We also observe microwave radiation strongly squeezed by 8 dB  below vacuum. A source of bright squeezed microwaves opens up applications in manipulations of quantum systems, and noiseless amplification can be used even at modest cryogenic temperatures.
\end{abstract}
\maketitle

The sensitive measurement of electromagnetic waves is instrumental in science and technology. A sinusoidally oscillating field $X(t) = X_1 \cos(\omega t) + X_2 \sin(\omega t)$ at the frequency $\omega$ is characterized by the quadrature amplitudes $X_1$ and $X_2$. In quantum mechanics, the quadratures are non-commuting observables which cannot be measured simultaneously. In a usual measurement which responds equally to both quadratures, noise must therefore increase by at least half the zero point fluctuations \cite{Caves,GirvinReview}.

In a phase-sensitive measurement, the two quadratures are amplified at different gain factors ${\cal G}_1$ and ${\cal G}_2$, such that the output quadratures are $Y_{i}={\cal G}_i X_i$. If either of the gains becomes very small and thus  the information in this quadrature is discarded, the other quadrature can be perfectly measured.
At the same time, the fluctuations in the discarded quadrature can become squeezed below the zero-point fluctuation level. Here we demonstrate such a nearly perfect measurement, proposed very recently \cite{SqAmpTheory}, of microwave light using a cavity optomechanical setup.
Along with the practical device whose design simplicity  shows promise for technological applications, we realize the phase mixing amplifier concept introduced in Ref.~\cite{SqAmpTheory}, and evolve the concept further.



The most sensitive measurements of microwave fields so far have taken advantage of nonlinearities of Josephson junctions \cite{Yurke1988Sq,Bergeal:2010iua,Lehnert2008Amp,Wallraff2modent,Siddiqi2015Amp}. Following the pioneering work in late 1980's \cite{Yurke1990Squ}, Josephson junction parametric amplifiers have reached the impressive system noise performance of 0.62 added quanta of noise  in the phase insensitive mode, close to the fundamental limit, and 0.14 quanta in the phase sensitive mode \cite{Deppe2013squ}. Therefore,  after almost three decades of development, these amplifiers are currently actively used in quantum science. Also electromechanical systems have recently been investigated for sensitive detection of electromagnetic signals \cite{MechAmpPaper,Bowen2012,Polzik2014Amp}. They usually exhibit a very narrow bandwidth, but recent work \cite{CasparAmp,KippenbergAmp} introduces a notable exception. Refs.~\cite{CasparAmp,KippenbergAmp} demonstrate a phase insensitive amplifier with a noise relatively low but not quite yet at the quantum limit.



%
\begin{figure*}
\centering
\includegraphics[width=0.95\textwidth]{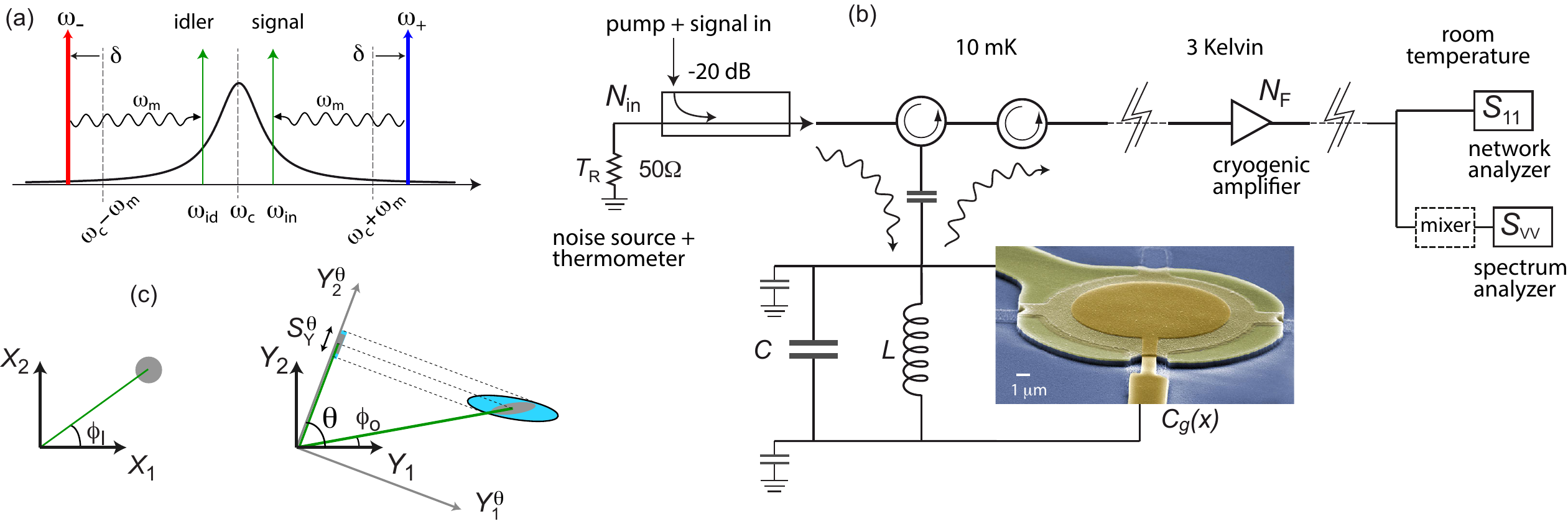}
\caption{\emph{Optomechanical phase-sensitive measurement.} (a), Two strong almost equal-amplitude pump tones (red and blue) are applied detuned from the sideband resonance $\omega_c \pm \omega_m$ by the amount $\delta \lesssim \kappa$. (b), Drumhead-type mechanical oscillator is coupled to a superconducting microwave-frequency ($LC$) resonator through a radiation-pressure force. A $50 \, \Omega$ resistor is used as an adjustable noise source with a controlled temperature $T_R$, directly connected through a directional coupler. The  cavity frequency is $\omega_c/2 \pi \simeq 6.9148$ GHz, cavity linewidth $\kappa/2 \pi \simeq 6.44$ MHz, internal cavity losses $\kappa_{\rm I}/2 \pi \simeq 90$ kHz, frequency of the mechanical mode $\omega_m/2 \pi \simeq 10.319$ MHz, and its linewidth $\gamma /2 \pi  \simeq 107$ Hz.  (c), A sinusoidal signal (line) is represented with the quadrature amplitudes $X_1$ and $X_2$. Before amplification (left), the gray circle denotes the quantum noise of the input signal. Following phase-sensitive amplification (right), the input quantum noise is squeezed into an ellipse. The total noise at the output has a contribution from the noise added in the process (light blue, ideally none). The signal plus noise is measured in a local oscillator basis defined by the angle $\theta$. }
\label{Fig1}
\end{figure*}

Our realization of a practically noiseless amplifier can be pictured as a generic cavity optomechanical setup. It  is physically simple, consisting of a superconducting microwave resonator, the cavity, with frequency $\omega_{c}$, coupled to a 15 microns wide membrane \cite{Teufel2011a} vibrating at the frequency $\omega_{m}$, as seen in \fref{Fig1}b. The two systems are coupled via the radiation pressure coupling $H_{\rm int}= g_0 n_{\rm c}\left(b^\dagger+b \right)$, where $n_{c} =  a^\dg  a$ is the number of microwave cavity photons, $ x =  b^\dagger +  b $ is the (dimensionless) position operator of the mechanical oscillator, and $g_0$ is a coupling constant. The cavity and the oscillator have the respective decay rates $\kappa$ and $\gamma$. The cavity is driven by two strong microwave tones of frequencies $\omega_+ = \omega_{c} + \omega_{m} + \delta $ and $\omega_- = \omega_{\rm c} -\omega_{m} -\delta$ (\fref{Fig1}a). The pumps induce respective cavity fields of amplitude $\alpha_+$ and $\alpha_-$. \vari{Here, $\delta$ describes the detuning from the exact blue/red sideband co-resonance condition.
The pumping results in an enhanced linear coupling of strength $G_\pm = g_0 \alpha_\pm$.}

This pump scheme is related to back-action evading measurements \cite{Schwab2014QND} and squeezing  \cite{SchwabSqueeze,Squeeze,TeufelSqueeze} of the mechanical oscillator, and to the dissipative squeezing recently proposed in Ref.~\cite{Clerk2014DissipSqu}. However, introducing a detuning $\delta \gtrsim (G_-^2-G_+^2)/\kappa$ drastically changes the resulting physics. In the following, we suppose operation in the resolved-sideband regime, where $\omega_m \gg \kappa$.  As shown in \cite{supplement}, the resulting Hamiltonian describing this system is
\begin{equation}
\label{hamilt}
H = \delta  b^\dagger  b + G_+ ( a^\dagger  b^\dagger +  b  a ) + G_- ( a^\dagger  b +  b^\dagger  a) \,.
\end{equation}
%
We make a Bogoliubov transformation of the cavity to a set of new operators,  $ \alpha$ so that $ a=u  \alpha - v \alpha^\dagger$. We choose $u=\cosh(\xi), \, v=\sinh(\xi)$ with the real parameter $\xi$ satisfying $\tanh(\xi)=G_+/G_-$. The resulting cavity-oscillator Hamiltonian is that of a beam splitter with coupling strength $G_{BG}=\sqrt{G_-^2-G_+^2}$, known to lead to the cooling of the mechanical oscillator \cite{Teufel2011a}. If the cavity is overcoupled, signals sent to it are completely reflected, i.e., the reflected signal at a given frequency $\omega$ experiences only a phase shift, or $\alpha_{\rm out}(\omega)=e^{i\phi(\omega)} \alpha_{\rm in}(\omega)$. In a large range of frequencies determining the amplifier bandwidth, different phase shift of the Bogoliubov wave at positive and negative frequencies translates into phase-sensitive or phase-mixing amplification in the original cavity frame \cite{supplement}. Even though the mechanical oscillator in a typical situation resides at a high thermal occupation number $n^T_m \simeq k_b T/ \hbar \omega_m \gg 1$ the added noise related to the input is of the order of $\LL( \gamma \kappa/G_{BG}^2 \RR) n_m^T$ and can be almost neglected in our setup due to the low value of $\gamma \ll G_{BG}^2/\kappa$. One can thus reach very nearly quantum limited operation even when the mechanical oscillator is far from its quantum ground state.

\begin{figure*}[t]
\centering
\includegraphics[width=0.95\textwidth]{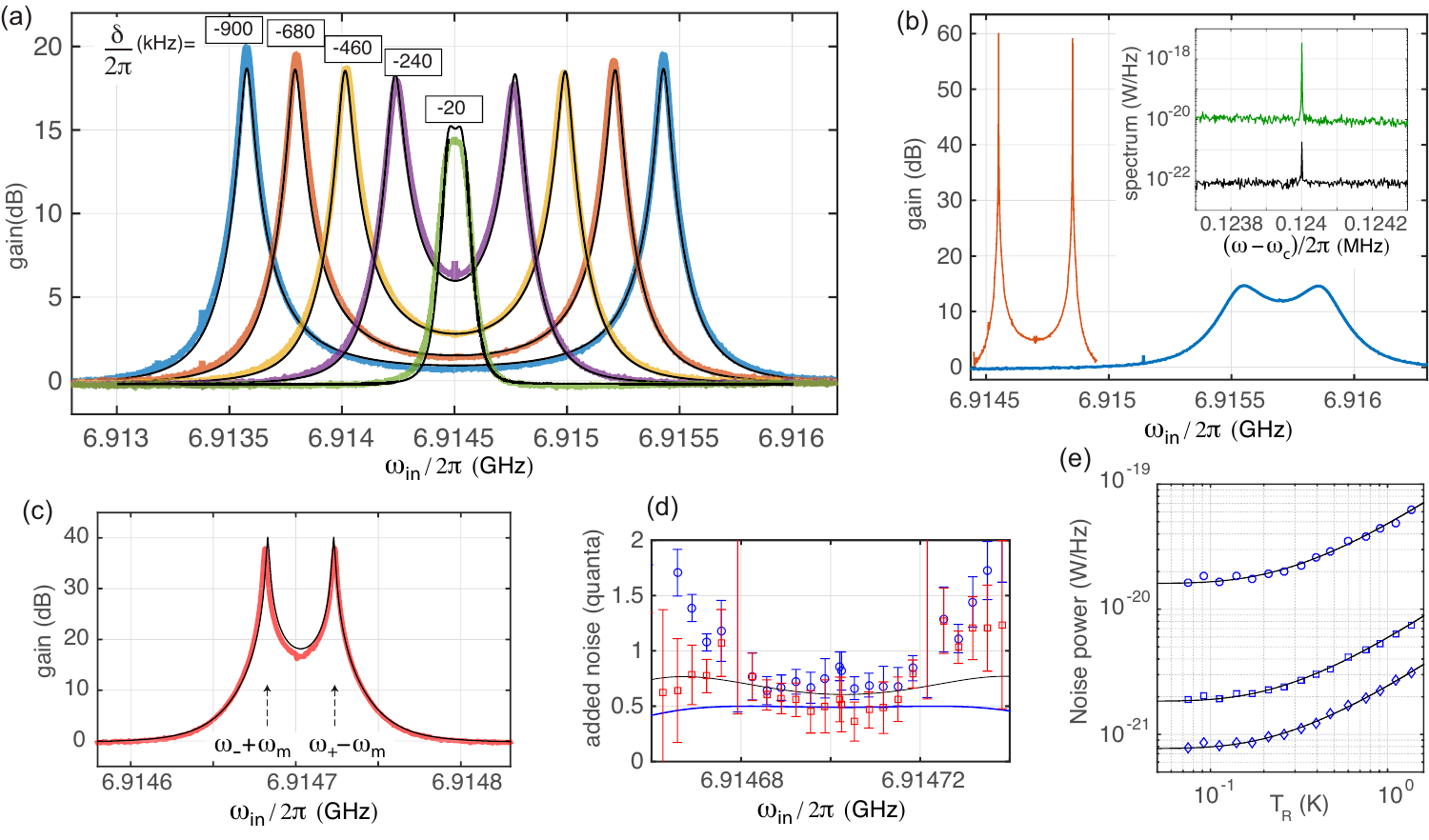}
\caption{\emph{Phase-insensitive amplification and noise.} (a), Gain at a fixed pump power, and at varying pump detunings. Black curves are theory predictions with $G_-/2\pi = 580$ kHz, $G_+/2\pi = 496$ kHz. (b), Amplification at a high gain (red), and over a broad bandwidth (blue). Inset: Improvement of the signal-to-noise ratio of a coherent input signal (sharp peak). The original (black) noise floor is limited by the commercial cryogenic amplifier. When the pump tones are switched on (green), the signal-to-noise ratio in phase-insensitive amplification is improved by 12 dB. (c), Gain for $G_-/2\pi \simeq 308$ kHz, and $\delta/2\pi  \simeq 20$ kHz. The black line is a theory prediction.  (d), Added noise corresponding to panel c: Blue circles are the total system noise, and red squares represent the added noise due to the optomechanical amplification. The solid black line is a theory curve, while the blue line shows the quantum limit. (e), Noise calibration by varying the power emitted by a known noise source. The noise at device output is shown. The data sets correspond to $\omega_{in}/2\pi = 6.914682,  6.914686$, and $6.914689$ GHz, in (c), from top to bottom. The solid lines are fits to the quantum noise formula.}
\label{totalgain}
\end{figure*}

At this point, let us discuss a generic phase-sensitive amplifier \cite{Caves}, referring to \fref{Fig1}c. At the input on top of a coherent signal, there is quantum noise, which usually does not show a phase preference. Hence the possible values of the  quadrature amplitudes $X_1$ and  $X_2$ of the input signal $X$ fall uniformly inside the gray circle representing the variance. Following phase-sensitive amplification, the input noise gets squeezed into an ellipse owing to unequal gains $\mathcal{G}_1$ and  $\mathcal{G}_2$ for the input quadratures. The principal axes in the amplified input noise define the preferred (output) quadratures which obey $Y_1= \mathcal{G}_1 X_1$, $Y_2= \mathcal{G}_2  X_2$, and the average amplified signal is $Y^2 = 1/2 \LL( Y_1^2 + Y_2^2 \RR) = \mathcal{G}^2 X^2$, with the total gain $\mathcal{G}$. In the amplification, some noise $N_{\m{add}}$ is added. In the present case, corresponding to phase-mixing amplification \cite{SqAmpTheory},  the added-noise ellipse is rotated with respect to the  input noise ellipse. 

Phase-sensitive amplification requires specifying a carrier frequency around which the quadrature operators are defined \cite{Caves}. In our setup, the carrier frequency is the center frequency of the pumps, $\omega_{0} \equiv (\omega_- + \omega_+)/2$. The carrier frequency not only defines the output quadratures, but the input (preferred) quadratures as well. Therefore, unless the input signal lies exactly at $\omega_{0}$, a rigorous definition of the input quadratures requires the presence of two fields symmetrically centered around $\omega_0$. The latter case means that one has to consider a field also at the idler (or, image) frequency $\omega_{id}$, satisfying $2 \omega_0 = \omega_{in} + \omega_{id}$, as illustrated in \fref{Fig1}a. In that case it is the idler that adds at least the half quantum of noise. 


Apart from the preferred basis, the amplified signal and the added noise can be read in a frame defined by a local oscillator (LO) having a phase $\theta$. In \fref{Fig1}c, the latter is indicated by the projections on the $Y_2^\theta$ axis. We can hence define a $\theta$ dependent gain $\mathcal{G}_\theta$ whose precise form depends on the phase of the input signal. In contrast to earlier work, \vari{e.g.~Refs.~\cite{Yurke1988Sq,Lehnert2008Amp,Deppe2013squ},} we treat the general case  where $\theta$ is not along the preferred basis. This scheme is the phase mixing amplifier \cite{SqAmpTheory}, which can provide an improved signal-to-noise because the added noise can have a nontrivial dependence on $\theta$. The added noise properties are characterized by referring the noise to the input, that is, by the spectral density $S_X^\theta = S_{Y}^\theta/\mathcal{G}^2_\theta$ where $S_{Y}^\theta$ represents the output spectral noise when no input signal is present. Expressed in units of quanta at the signal frequency, the phase-dependent added noise is $N_{\m{add}}^\theta = S_X^\theta/ \hbar \omega_{in}$. 


%
\begin{figure}[t]
\includegraphics[width=0.4\textwidth]{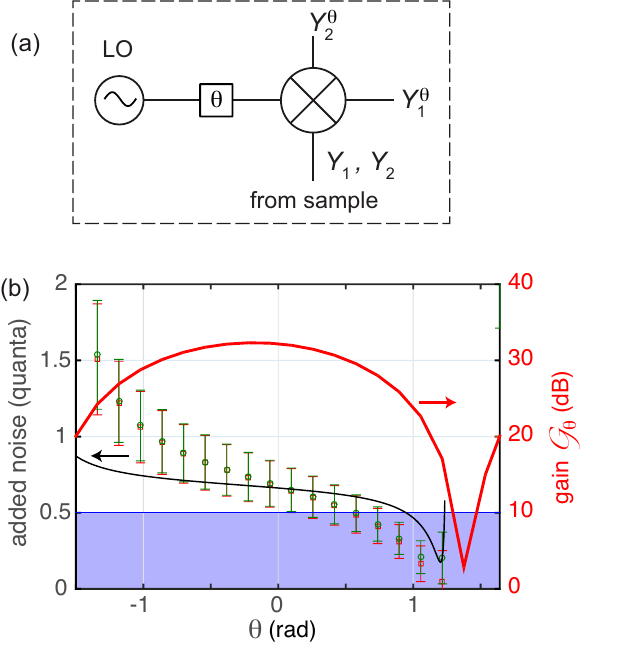}
\caption{\emph{Phase-sensitively measured noise.} (a), In homodyne detection, a mixer (dashed box in \fref{Fig1}b) extracts the quadrature amplitudes oscillating at the local oscillator (LO) frequency $\omega_{\m{LO}}$. The quadrature axes $Y_1^\theta$ and $Y_2^\theta$ are determined by the digitally tunable phase shift $\theta$. (b), Added input noise $N_{\m{add}}^\theta$ (left scale) and gain (right scale) as a function of quadrature angle $\theta$. The green circles are the system noise, and the red squares represent the contribution by the electromechanical device. The regime below the quantum limit is colored in blue. The parameters are $G_-/2\pi \simeq 320$ kHz, $G_+/2\pi \simeq 315$ kHz, $\delta /2\pi  \simeq 460$ kHz, and  $\omega_{in}/2\pi = 6.915165$ GHz.} 
\label{phasesensit}
\end{figure}
%

We perform the experiments in a dilution refrigerator at a temperature of approximately 10 mK. 
The basic signal scheme is shown in \fref{Fig1}b. The two microwave pump tones and a weak signal tone are applied to the coupler port of the device. The amplification is measured with a network analyzer as the $S_{11}$ reflection parameter. In \fref{totalgain}a-c we demonstrate phase-insensitive amplification of microwaves achievable with the scheme.
The double-peak structure corresponds to the positions of the resonances of the signal and idler, that is, where a phonon in the mechanical oscillator is emitted or absorbed by a pump tone. As shown in \fref{totalgain}b, we observe high amplification up to 60 dB, as well as a broad 3 dB bandwidth ( $\simeq 430$ kHz). The data shown in \fref{totalgain}c correspond to the noise measurements in \fref{totalgain}d, discussed below. The theoretical predictions \cite{supplement}, overlaid on the experimental data, show a good agreement. In order to quantitatively explain the gain profiles, we include a parametric modulation term to the mechanical resonator as in Ref.~\cite{Schwab2012instab}.

For noise measurements, we use a $ 50 \, \Omega$ resistor as a tunable known noise source. It is attached to a heater and a separate thermometer, and connected to the sample via a short superconducting coaxial cable. At the known calibration temperature $T_R$, the quantum noise power from the resistor is $N_{\m{in}} = \coth(\hbar \omega_c/(2 k_b T_R))/2$.  This calibrated input noise gives rise to an output noise power of $P_{\m{RT}} = N_{\m{in}} G F  + \LL( N_{\m{add}} + N_F/G\RR) G F $ at room temperature. Here, $F$ and $N_F \simeq 18 \pm 2$ quanta  \cite{supplement} are, respectively, the gain and the technical noise due to all amplifiers and attenuation following the sample. We use the quantities $\LL( N_{\m{add}} + N_F/G\RR)$ and $GF$ as adjustable parameters when fitting data to the expression for $P_{\m{RT}}$ at varying values of $T_R$. In \fref{totalgain}e we display an example of the measured power as function of $T_R$, showing a good agreement with the expected quantum noise. 

The total  (averaged over quadratures) noise corresponds to a phase-insensitive measurement, with the measurement frequency different from $\omega_0$. As shown in \fref{totalgain}d, we observe a total system noise well below the single quantum level, and the added noise $N_{\m{add}}$ is consistent with the quantum limit of 0.5 quanta. The theory curves include dielectric heating of the baths by the pumps up to $n_m^T \simeq 80$, $n_I \simeq 1.1$. \vari{Here, $n_I$ is photon occupation of the internal bath of the cavity mode. In a previous cooldown, we made a rough independent calibration of the bath heating by using sideband cooling, observing a sharp onset of heating around the powers discussed here  \cite{supplement}. The low noise appears clearly as an improvement of signal-to-noise ratio of a weak signal when the amplification is switched on, as displayed in \fref{totalgain}b (inset), where the observed noise floor is dominated by amplified vacuum noise. Figure \ref{totalgain}d shows that the added noise can be very close to 0.5 quanta even when the mechanical oscillator responsible for the amplification resides far from its ground state. 

For phase-sensitive homodyne measurements, the output signal is digitally mixed to the center frequency $\omega_{\m{LO}}=\omega_{0} $ in the scheme of \fref{phasesensit}a. By changing the phase $\theta$ of the LO, we obtain the quadratures $Y_1^\theta(t)$ and  $Y_2^\theta(t)=Y_1^{\theta + \pi/2}(t)$ oscillating at the center frequency. 
The noise measurements are made as described above, but individually for each  $\theta$ \cite{supplement}. We discover that the noise falls well below the quantum limit in one quadrature (\fref{phasesensit}b), and we estimate $N_{\m{add}} \lesssim 0.2$ quanta. This uncertainty is mostly limited by the statistical errors from fitting to the quantum noise. The theoretical curve (black) in \fref{phasesensit}b is evaluated using $n_m^T \simeq 300$, $n_I \simeq 1.5$, capturing the main features involving the optimum noise detuned from the preferred quadrature. 

%
\begin{figure}
\includegraphics[width=0.5\textwidth]{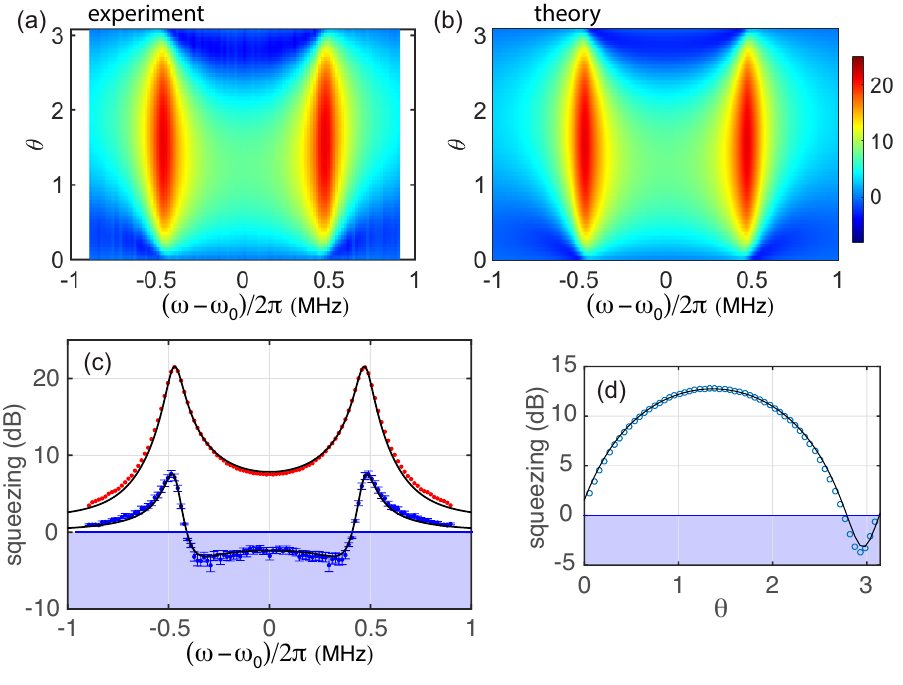}
\caption{\emph{Mechanical squeezing of microwave light.} (a), Noise emitted from the device normalized to the vacuum level (in dB), as a function of the local oscillator phase $\theta$ and frequency. (b), Theoretical prediction corresponding to (a).  (c), Cross-cut along the horizontal lines $\theta \simeq 2.93$ (blue), and $\theta = 1.38$ (red). The thin solid lines are the corresponding theoretical curves. (d), Squeezing at $\omega/2\pi = -0.369$ MHz as a function of $\theta$. In (c) and  (d), the area under the zero-point fluctuation level is colored in blue. The pump amplitudes are $G_-/2\pi \simeq 690$ kHz, $G_+/2\pi \simeq 590$ kHz.}
\label{squeeze}
\end{figure}
%


A fundamental property of a phase-sensitive amplifier is the possibility to generate squeezed propagating states  as shown in many experiments in optics \cite{Squeeze1985,Kimble1986Squ}, and with Josephson devices, see e.g.~\cite{Yurke1990Squ,Lehnert2008Amp}. Quantum squeezing of the light emitted from optomechanical cavities has also recently been observed at optical frequencies \cite{Atom2012Sq,Painter2013Sq,Regal2013Sq}. Next we show that our approach provides a way to generate squeezed radiation. This demonstrates a new mechanism of  over the previously utilized ponderomotive squeezing \cite{Atom2012Sq,Painter2013Sq,Regal2013Sq}. In the plane of the input of the cryogenic amplifier, we measure strong squeezing within a bandwidth of 700 kHz, with the maximum of $\simeq 3.5$ dB below vacuum, as shown in \fref{squeeze}. The calibration procedure is described in the supplementary \cite{supplement}. The theory predictions in \fref{squeeze}b-d are generated using $n_m^T = 400$, $n_I = 1.6$. We infer that the amount of squeezing, depleted by losses before the cryogenic amplifier, right following the sample has been up to 8 dB. This value is on par to those obtained with Josephson parametric amplifiers (JPA), e.g.~10 dB in Ref.~\cite{Lehnert2008Amp}.



Intense squeezed coherent states are a valuable resource \cite{Polzik1998Teleport,Zoller2009Squ,Siddiqi2013Squ,Zoller2009Squ,Teufel2016Squ}.  When injected with a sinusoidal signal, we estimate the setup of \fref{squeeze}  to produce a bright squeezed coherent state of up to $\sim 10^{14}$ photons/sec, or $ \sim -65$ dBm.
If realized in optics, our approach can provide luminous squeezed laser beams to  overcome the quantum-noise limitation  in gravitational wave observations  in particular because the squeezing angle is naturally frequency dependent as seen in \fref{squeeze}a,b.
Our  scheme of signal amplification compares favorably over JPA in the sense that it does not require superconductivity, and is able to handle 4 orders of magnitude more input power than a corresponding JPA \cite{Lehnert2008Amp}, or 2 orders of magnitude more than in Ref.~\cite{Siddiqi2015Amp}. Moreover, in contrast to cavity-based parametric amplifiers, the gain-bandwidth product is unlimited \cite{supplement}.  The instantaneous bandwidth in the current design is smaller than in JPA, but it can be increased by stronger coupling, or implementing an electromechanical metamaterial. With slight improvements, the device can operate below the quantum limit at modest cryogenic temperatures of a few Kelvin, hence offering an attractive technology for narrow-band measurements. 


\emph{Acknowledgements} We would like to thank Visa Vesterinen, Pasi L\"ahteenm\"aki and Timo Hyart for useful discussions. This work was supported by the Academy of Finland (contract 250280, CoE LTQ, 275245) and by the European Research Council (240387-NEMSQED, 240362-Heattronics, 615755-CAVITYQPD). The work benefited from the facilities at the OtaNano - Micronova Nanofabrication Center and at the Low Temperature Laboratory. 

\bibliography{/Users/masillan/Documents/latex/MIKABIB}

\end{document}